\documentclass[conference,a4paper]{IEEEtran}
\IEEEoverridecommandlockouts

\usepackage{cite}
\usepackage{amsmath,amssymb,amsfonts}
\usepackage{graphicx}
\usepackage{textcomp}
\usepackage{booktabs}
\usepackage{xcolor}
\usepackage{subcaption}
\usepackage{url}
\usepackage{algorithm}
\usepackage{algpseudocode}
\usepackage[acronym]{glossaries}

\def\BibTeX{{\rm B\kern-.05em{\sc i\kern-.025em b}\kern-.08em
    T\kern-.1667em\lower.7ex\hbox{E}\kern-.125emX}}

\makeatletter
\def\bstctlcite{\@ifnextchar[{\@bstctlcite}{\@bstctlcite[@auxout]}}
\def\@bstctlcite[#1]#2{\@bsphack
 \@for\@citeb:=#2\do{%
   \edef\@citeb{\expandafter\@firstofone\@citeb}%
   \if@filesw\immediate\write\csname #1\endcsname{\string\citation{\@citeb}}\fi}%
 \@esphack}
\makeatother

\newacronym{IDS}{IDS}{Intrusion Detection System}
\newacronym{CAN}{CAN}{Controller Area Network}
\newacronym{HPC}{HPC}{Hardware Performance Counter}
\newacronym{DLC}{DLC}{Data Length Code}
\newacronym{DoS}{DoS}{Denial-of-Service}
\newacronym{FS}{FS}{Full System}
\newacronym{RTOS}{RTOS}{Real-Time Operating System}
\newacronym{AES}{AES}{Advanced Encryption Standard}
\newacronym{SVC}{SVC}{Support Vector Classification}
\newacronym{ECU}{ECU}{Electronic Control Unit}
\newacronym{SDV}{SDV}{Software Defined Vehicle}
\newacronym{LTE}{LTE}{Long-Term Evolution}
\newacronym{5G}{5G}{Fifth Generation Mobile Network}
\newacronym{LIN}{LIN}{Local Interconnect Network}
\newacronym{SENT}{SENT}{Single Edge Nibble Transmission}
\newacronym{ETH10TBASES1}{Ethernet 10TBASE-S1}{10 Mbps Single Pair Ethernet}
\newacronym{UNR155}{UNR 155}{United Nations Regulation No. 155}
\newacronym{UNR156}{UNR 156}{United Nations Regulation No. 156}
\newacronym{CSMS}{CSMS}{Cybersecurity Management System}
\newacronym{ISO21434}{ISO 21434}{ISO/SAE 21434 Road Vehicles – Cybersecurity Engineering}
\newacronym{cybersecurity}{Cybersecurity}{Cybersecurity}
\newacronym{OTA}{OTA}{Over-the-Air}
\newacronym{FOTA}{FOTA}{Firmware Over-the-Air}
\newacronym{ADAS}{ADAS}{Advanced Driver Assistance Systems}
\newacronym{V2X}{V2X}{Vehicle-to-Everything}
\newacronym{AI}{AI}{Artificial Intelligence}
\newacronym{ML}{ML}{Machine Learning}
\newacronym{OEM}{OEM}{Original Equipment Manufacturer}
\newacronym{ISO24089}{ISO 24089}{Road vehicles — Software update engineering}
\newacronym{CNN}{CNN}{Convolutional Neural Network}
\newacronym{LSTM}{LSTM}{Long Short-Term Memory}
\newacronym{BERT}{BERT}{Bidirectional Encoder Representations from Transformers}
\newacronym{ID}{ID}{Identifier}
\newacronym{SVM}{SVM}{Support Vector Machine}
\newacronym{KNN}{KNN}{K-Nearest Neighbors}
\newacronym{DT}{DT}{Decision Tree}
\newacronym{HIDS}{HIDS}{Host-Based Intrusion Detection System}
\newacronym{AUTOSAR}{AUTOSAR}{Automotive Open System Architecture}
\newacronym{SOC}{SOC}{Security Operations Center}
\newacronym{TPM}{TPM}{Trusted Platform Module}
\newacronym{HSM}{HSM}{Hardware Security Module}
\newacronym{PMU}{PMU}{Perfomance Monitoring Unit}
\newacronym{RBF}{RBF}{Radial Basis Function}
\newacronym{TPR}{TPR}{True Positive Rate}
\newacronym{FNR}{FNR}{False Negative Rate}
\newacronym{FPR}{TPR}{False Positive Rate}
\newacronym{DNN}{DNN}{Deep Neural Network}

\begin{document}

\title{CANDoSA: A Hardware Performance Counter-Based Intrusion Detection System for DoS Attacks on Automotive CAN bus\\
\thanks{This work was supported by Project SERICS through the MUR National Recovery and Resilience Plan, funded by the European Union NextGenerationEU under Grant PE00000014, project COLTRANE-V funded by the Ministero dell'Università e della Ricerca within the PRIN 2022 program (D.D.104 - 02/02/2022), and  Project Vitamin-V funded by the European Union: Project 101093062.}
}

\author{\IEEEauthorblockN{Franco Oberti}
\IEEEauthorblockA{\textit{Dumarey Softronix} \\
\textit{Dumarey}\\
Torino, Italy \\
ORCID: 0000-0001-7974-9505}
\and
\IEEEauthorblockN{Stefano Di Carlo}
\IEEEauthorblockA{\textit{Control and Computer Eng. Dep.} \\
\textit{Politecnico di Torino}\\
Torino, Italy \\
ORCID: 0000-0002-7512-5356}
\and
\IEEEauthorblockN{Alessandro Savino}
\IEEEauthorblockA{\textit{Control and Computer Eng. Dep.} \\
\textit{Politecnico di Torino}\\
Torino, Italy \\
ORCID: 0000-0003-0529-7950}
}


\maketitle


\begin{abstract}
The \gls{CAN} protocol, essential for automotive embedded systems, lacks inherent security features, making it vulnerable to cyber threats, especially with the rise of autonomous vehicles. Traditional security measures offer limited protection, such as payload encryption and message authentication. This paper presents a novel \gls{IDS} designed for the \gls{CAN} environment, utilizing \glspl{HPC} to detect anomalies indicative of cyber attacks. A RISC-V-based \gls{CAN} receiver is simulated using the gem5 simulator, processing \gls{CAN} frame payloads with AES-128 encryption as FreeRTOS tasks, which trigger distinct \gls{HPC} responses. Key \gls{HPC} features are optimized through data extraction and correlation analysis to enhance classification efficiency. Results indicate that this approach could significantly improve \gls{CAN} security and address emerging challenges in automotive cybersecurity.
\end{abstract}


\glsresetall

\section{Introduction}
\label{sec:introduction}

Modern vehicles are now interconnected platforms capable of semi-autonomous decision-making and \gls{OTA} updates. This integration of automotive engineering and information technology necessitates a rethinking of vehicle architectures to ensure functionality, performance, efficiency, and resilience.

Vehicle communication utilizes networks such as \gls{CAN}, \gls{LTE}, \gls{5G}, and Ethernet to enable advanced features like \gls{ADAS}, \gls{V2X}, and cloud-based services \cite{Brown:2022aa, Miller:2023aa}. However, this connectivity increases vulnerability to cyber threats  \cite{Williams:2023aa, Anderson:2023aa}, including vehicle hijacking, \glspl{ECU} manipulation, and ransomware attacks \cite{Davis:2024aa}. The shift to \gls{SDV} introduces additional risks such as \gls{DoS} attacks and data injection, making robust, multi-layered cybersecurity frameworks essential \cite{Nguyen:2024aa}.

Regulations like \gls{UNR155} and \gls{UNR156} require manufacturers to implement \glspl{CSMS} to ensure compliance \cite{Jones:2024aa}. Standards such as \gls{ISO21434} emphasize the need for intrusion detection to protect communication networks \cite{Schmidt:2024aa}.

A common cybersecurity approach includes deploying \glspl{IDS} that analyze network traffic in real-time to identify threats. These systems leverage signature-based detection, anomaly detection, and machine learning to monitor networks like \gls{CAN} and Ethernet \cite{Kim:2023aa}, while host-based \glspl{IDS} enhance security by tracking critical \glspl{ECU} modifications.

This paper presents research on a novel \gls{IDS} aimed at improving attack detection on \gls{CAN} networks by assessing \glspl{HPC} deviations in application execution. We use a RISC-V microprocessor as a representative architecture for next-generation automotive systems~\cite{Cuomo:2023aa}.

The paper is structured as follows: Section \ref{sec:background} provides background information, Section \ref{sec:hw_ids} describes the \gls{IDS} framework, Section \ref{sec:simulation} outlines the simulation environment, Section \ref{sec:results} discusses results, and Section \ref{sec:conclusion} concludes the paper.

\section{Background}
\label{sec:background}

Automotive \glspl{IDS} encompass signature-based, anomaly-based, and hybrid approaches for vehicle cybersecurity \cite{Luo2023}. Signature-based systems detect known threats using predefined patterns, while anomaly-based solutions leverage \gls{ML} and \gls{AI} to identify novel attacks through behavioral analysis \cite{US10630699B2}. Hybrid approaches combine these methods for comprehensive protection, though they face challenges in balancing detection accuracy with resource constraints.

Recent advances have integrated sophisticated \gls{ML} techniques, particularly for \gls{CAN} bus monitoring. The \gls{CAN}-BERT model \cite{Alkhatib:2022aa} applies language modeling to analyze network traffic patterns, while traditional classifiers like \glspl{SVM} and \glspl{DNN} detect various attacks, including \gls{DoS} and impersonation attempts \cite{Bari2023, AIBasedIDS}.

Vehicle security employs both network-based \glspl{IDS} and \glspl{HIDS}. Network solutions monitor communication channels, while \glspl{HIDS} protect individual \glspl{ECU} through power signature analysis and host hardening techniques \cite{PHIDIAS2024}. Commercial implementations combine these approaches with cloud computing and big data analytics for real-time threat detection \cite{CyberattacksEVs}, though high computational requirements can impact response times \cite{Luo2023}.

Key challenges include achieving high \gls{TPR} while maintaining low latency in resource-constrained environments. Future directions focus on integrating hardware security modules (\glspl{TPM}, \glspl{HSM}) with software-based detection \cite{CyberattacksEVs}, implementing federated learning for privacy-preserved model training \cite{Bari2023}, and developing standardized evaluation frameworks \cite{Bouchouia:2024aa}. Recent research also explores \glspl{HPC} event counting for detecting malware and microarchitectural attacks \cite{Magliano:2024aa, Kasap:2023aa}.

\section{Hardware-Based Intrusion Detection Framework}
\label{sec:hw_ids}

\begin{figure}[htb]
    \centering
    \includegraphics[width=0.9\linewidth]{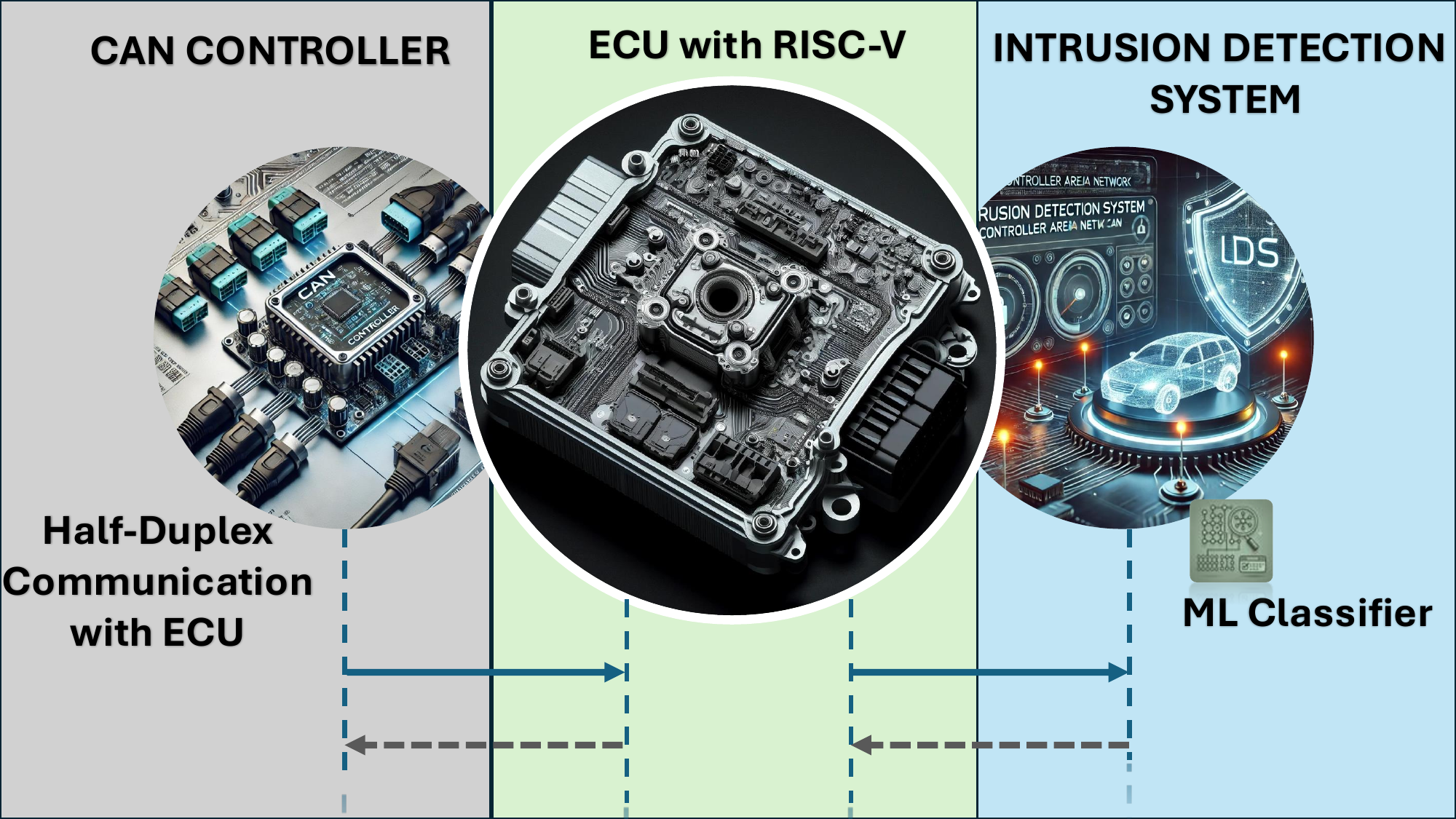}
    \caption{General workflow for attack detection.}
    \label{fig:global_workflow}
\end{figure}

This section outlines the framework of the \gls{IDS}, aimed at developing a \gls{HIDS} to identify attacks by detecting behavioral deviations in the application on the \gls{ECU} and analyzing \gls{CAN} data (see Figure \ref{fig:global_workflow}). The approach consists of three phases that transform raw \gls{CAN} data into an effective online inference model using \gls{HPC} event counting:

\begin{enumerate} 
\item \textbf{Data Collection}: Data is gathered from the \gls{CAN} network and organized into samples, which may be benign or malicious. This setup facilitates the detection of deviations in the \gls{ECU} 's behavior, signaling potential attacks. While it limits detectable scenarios, it effectively identifies intrusion events like \gls{DoS} attacks, which overwhelm ECU functionality, and frame spoofing attacks, where false messages are injected. 

\item \textbf{Frame Processing and \gls{HPC} Logging}: The \gls{CAN} data is processed by an application on the \gls{ECU}, activating the \gls{PMU} to log \gls{HPC} values. This logging captures events during frame processing, providing insights into the \gls{ECU} 's state and performance. Currently, \gls{HPC} data is collected only at execution's end, limiting real-time detection but remaining crucial for identifying performance deviations~\cite{Kuruvila:2022aa}. 

\item \textbf{Offline Training and Online Inference}: The classification model undergoes offline training, processing \gls{HPC} values through feature selection to identify key features and optimize classifier parameters. Once trained, the inference model continuously monitors \gls{CAN} data to detect attacks based on deviations in the \gls{ECU} 's behavior.
\end{enumerate}

\subsection{Classification Framework}
\label{subsec:classification}

The classification step is crucial for ensuring a fair analysis of \gls{HPC} data. Initially, the raw data undergoes preprocessing to standardize and normalize features, eliminating discrepancies from differing scales and filtering out irrelevant data. This process enhances comparability and lays a foundation for accurate analysis. Two key transformations are applied:

\begin{enumerate} 
\item \textbf{Mean and Scale Transformation:} This reduces dispersion in \gls{HPC} values, mitigating the impact of outliers from simulation timing or workload variations. 
\item \textbf{Correlation Analysis for Feature Reduction:} This analysis identifies and removes highly correlated features, reducing dimensionality while retaining the most predictive variables. The refined dataset focuses on \gls{HPC} features that distinguish normal from anomalous traffic.
\end{enumerate}

The classification employs Binary Classification Models to differentiate regular traffic from attacks. We utilize an unsupervised One-Class Classifier trained solely on standard traffic data, which requires less training data and effectively detects anomalies by identifying deviations from the established dataset. This approach ensures robust detection without constant updates, enhancing our cybersecurity efforts.

\section{Simulation of ECU-CAN systems}
\label{sec:simulation}

To deploy a realistic \gls{ECU} configuration, a \gls{CAN} receiver integrated with a RISC-V architecture is implemented on the gem5 simulator. The setup is configured using \texttt{riscv/fs\_linux.py}, which establishes a RISC-V Timing Simple CPU, DDR4 RAM, L1/L2 caches, and a 1.0 GHz clocked system, reflecting a typical embedded system. This controlled environment overcomes physical board limitations, enabling precise observation and manipulation of \gls{CAN} communication and \gls{HPC} triggering. The simulation runs in gem5's \gls{FS} mode, emulating critical hardware components and supporting complex software like Linux or \gls{RTOS} (e.g., FreeRTOS~\cite{Doc_FreeRTOS}).

\gls{CAN} communication involves two components: 

\begin{itemize} 

\item \textbf{\gls{CAN} Controller Transmitter:} A Python script reads datasets, constructs \gls{CAN} frames, and writes them to a shared file. Frames are generated by parsing components like identifiers, \gls{DLC}, and payloads and converting them into frames. 

\item \textbf{\gls{CAN} Controller Receiver:} A \emph{FreeRTOS} task reads and processes frame data from shared files, simulating memory-mapped access in the absence of a complete hardware counterpart. 
\end{itemize}

Since the gem5 RISC-V microprocessor lacks a \gls{PMU}, \glspl{HPC}-like data are derived from gem5 logs~\cite{Dutto2021}. The simulator tracks around 1200 architectural events in \texttt{stats.txt}, offering a broader dataset than physical \gls{PMU} counters. A Python script parses these logs to extract and reshape \gls{HPC}-like data for training and testing detection models, following preprocessing steps outlined in Section \ref{subsec:classification}.

\section{Experimental Results}
\label{sec:results}

\begin{figure}[tb]
    \centering
    \includegraphics[width=0.99\linewidth]{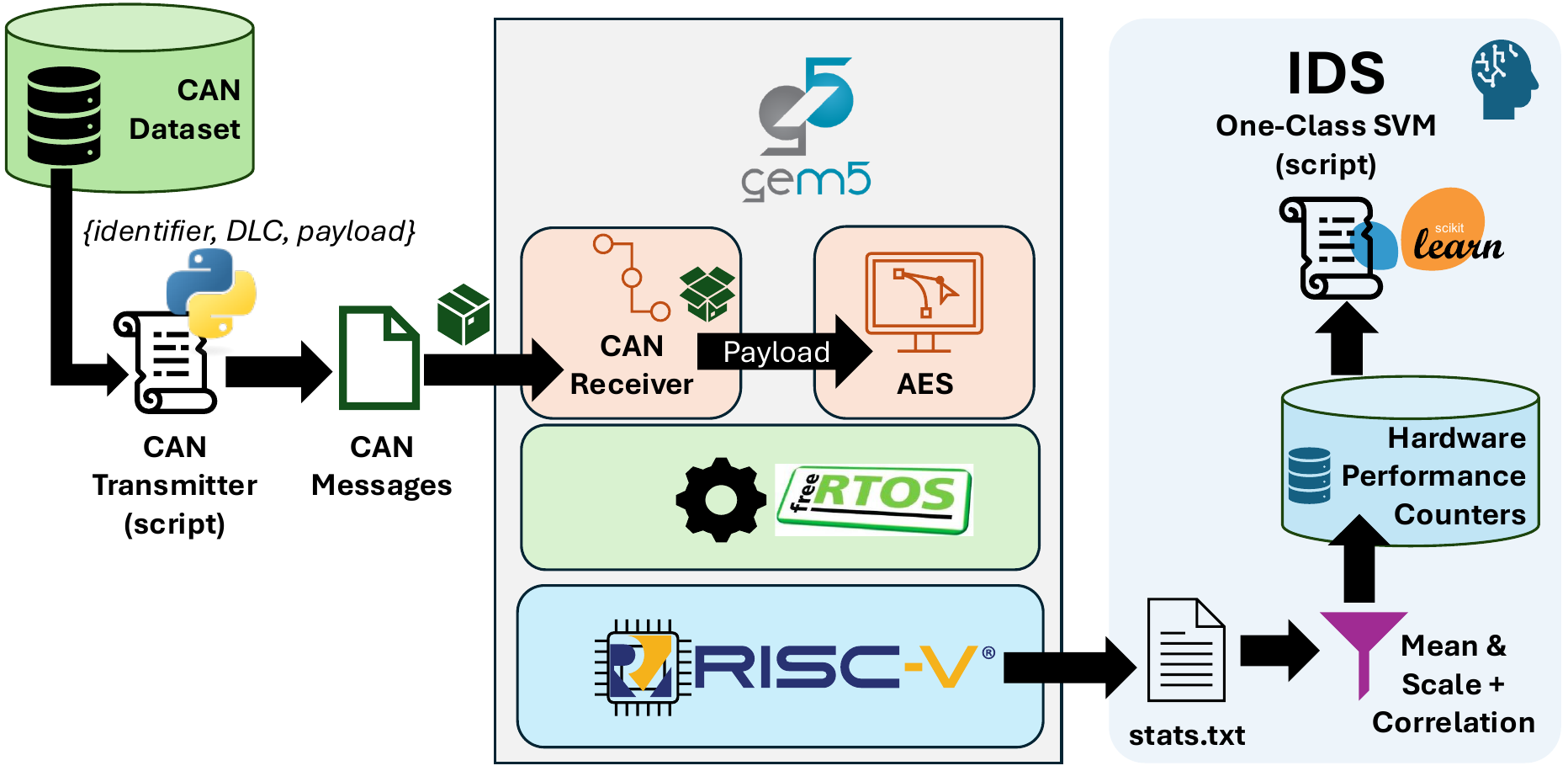}
    \caption{gem5 \cite{Doc_gem5} experimental environment}
    \label{fig:sim}
\end{figure}

The experimental setup, shown in Figure \ref{fig:sim}, features an \texttt{AES-128} encryption task as a reference application. This task processes 16-byte plaintext blocks with a 16-byte key to produce ciphertext. It runs as a \emph{FreeRTOS} task, receiving payloads as \gls{CAN} frames from the \gls{CAN} receiver described in \ref{sec:simulation}. This setup simulates a realistic automotive environment by combining \emph{FreeRTOS} and \gls{HPC} evolution beyond the \texttt{AES} algorithm.

\begin{figure}[htb]
\centering
\includegraphics[width=0.45\textwidth]{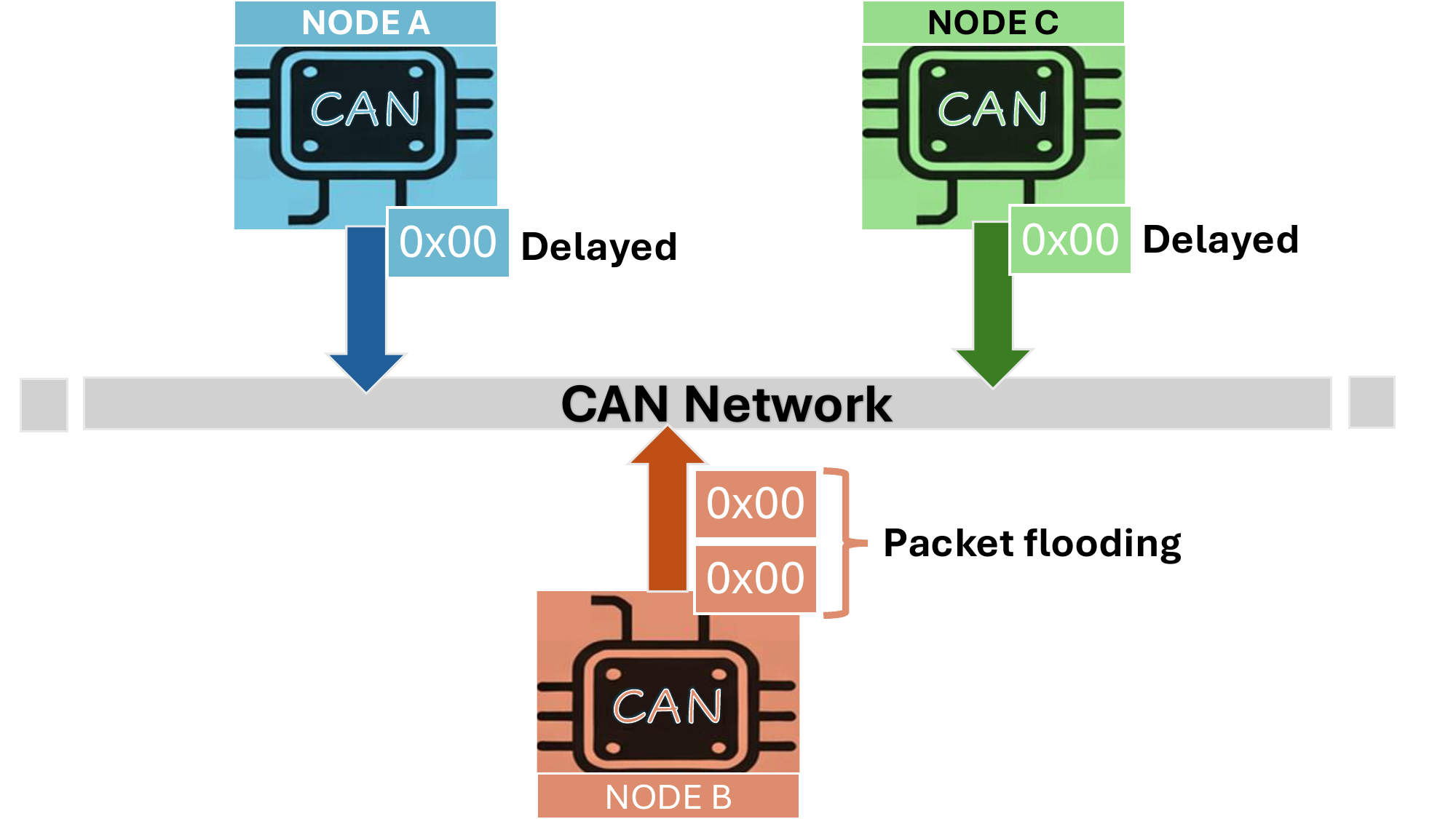}
\caption{Schematic of the DoS attack}
\label{fig:DoS_attack}
\end{figure}

\begin{table*}[htb]
\centering
\caption{Selected Events from gem5 logs (all start from \texttt{system.})}
\label{tab:hpc}
\resizebox{\textwidth}{!}{%
    \begin{tabular}{llcc}
        \toprule
        \textbf{gem5 Event} & \textbf{gem5 Meaning} & \textbf{RISC-V \gls{HPC} Similarity} & \textbf{x86 \gls{HPC} Similarity~\cite{intel_sdm}} \\
        \midrule
        cpu.commitStats0.numInsts       & Committed instructions                           & \texttt{minstret} (Retired instruction counter) \cite{riscv_spec}   & INST\_RETIRED.ANY \\
        \addlinespace
        cpu.fetchStats0.numBranches     & Fetched branch instructions                      & Branch instructions event (PULP) \cite{pulp_doc}                  & BR\_INST\_RETIRED.ALL\_BRANCHES \\
        \addlinespace
        cpu.dcache.demandHits::cpu.data  & Demand hits in the data cache                    & L1 D-Cache hit event (PULP) \cite{pulp_doc}                        & MEM\_LOAD\_RETIRED.L1\_HIT \\
        \addlinespace
        cpu.dcache.demandMisses::cpu.data& Demand misses in the data cache                  & L1 D-Cache miss event (PULP) \cite{pulp_doc}                       & MEM\_LOAD\_RETIRED.L1\_MISS \\
        \addlinespace
        cpu.dcache.ReadReq.hits::cpu.data& Read request hits in the data cache              & Load access event (CVA6) \cite{cva6_doc}                           & MEM\_LOAD\_RETIRED.L1\_HIT (subset) \\
        \addlinespace
        cpu.dcache.ReadReq.misses::cpu.data& Read request misses in the data cache           & Load access event (CVA6) \cite{cva6_doc}                           & MEM\_LOAD\_RETIRED.L1\_MISS (subset) \\
        \addlinespace
        cpu.dcache.WriteReq.hits::cpu.data& Write request hits in the data cache             & Store access event (CVA6) \cite{cva6_doc}                          & MEM\_STORE\_RETIRED.L1\_HIT \\
        \addlinespace
        cpu.dcache.WriteReq.misses::cpu.data& Write request misses in the data cache           & Store access event (CVA6) \cite{cva6_doc}                          & MEM\_STORE\_RETIRED.L1\_MISS \\
        \addlinespace
        cpu.icache.demandHits::cpu.inst  & Demand hits in the instruction cache             & L1 I-Cache hit event (PULP) \cite{pulp_doc}                         & ICACHE.HIT \\
        \addlinespace
        cpu.icache.demandMisses::cpu.inst& Demand misses in the instruction cache           & L1 I-Cache miss event (PULP) \cite{pulp_doc}                        & ICACHE.MISS \\
        \addlinespace
        cpu.icache.ReadReq.hits::cpu.inst& Read request hits in the instruction cache         & Instruction fetch event (CVA6) \cite{cva6_doc}                     & ICACHE.HIT (subset) \\
        \addlinespace
        cpu.icache.ReadReq.misses::cpu.inst& Read request misses in the instruction cache      & Instruction fetch event (CVA6) \cite{cva6_doc}                     & ICACHE.MISS (subset) \\
        \addlinespace
        l2.demandHits::cpu.data          & Demand hits in the L2 cache                      & L2 cache hit event (if implemented in PULP/CVA6)                     & MEM\_LOAD\_RETIRED.L2\_HIT \\
        \addlinespace
        l2.demandMisses::cpu.inst        & Demand misses in the L2 cache (instructions)     & L2 cache miss event (if implemented in PULP/CVA6)                    & MEM\_LOAD\_RETIRED.L2\_MISS (for instructions) \\
        \addlinespace
        l2.demandMisses::cpu.data        & Demand misses in the L2 cache (data)             & L2 cache miss event (if implemented in PULP/CVA6)                    & MEM\_LOAD\_RETIRED.L2\_MISS (for data) \\
        \addlinespace
        l2.demandMisses::total           & Total demand misses in the L2 cache              & Total L2 cache miss event (if implemented in PULP/CVA6)                & L2\_RQSTS.MISS \\
        \bottomrule
    \end{tabular}%
}
\end{table*}

The dataset, sourced from \cite{Hyunsung2017aa}, was collected via the OBD-II port of a KIA SOUL car and includes regular (attack-free) data and three attack types: Fuzzy, Impersonation, and \gls{DoS}. For this study, we focused on the \gls{DoS} attack, where a malicious node injects \gls{CAN} frames with an identifier of zero, causing arbitration delays for legitimate frames. The dataset contains 119,000 attack-free messages and 118,000 \gls{DoS} attack messages.To evaluate the \gls{IDS} framework, we used the unsupervised One-Class SVM~\cite{Mahfouz2020}, implemented via \texttt{sklearn}~\cite{Doc_sklearn}. This model, trained on attack-free data, identifies anomalies by establishing a boundary around the normal class. Training utilized 20\% to 95\% of the attack-free data, with the remaining 5\% reserved for testing alongside the full attack dataset.

After parameter tuning, the \texttt{OneClassSVM} was configured with an \gls{RBF} kernel, $\nu = 0.2$, and $\gamma = auto$. This setup effectively detected anomalies within the dataset.

We implemented the two methods outlined in Section \ref{subsec:classification} to streamline our data. We standardized the input values, which ranged from \(10^{-7}\) to \(10^{10}\), to achieve a mean of zero and a standard deviation of one, ensuring uniformity in our analysis. We computed correlation coefficients for each input parameter with the output parameter, excluding any that were uncalculable or did not meet the 0.9 threshold. This procedure ultimately informed the final selection in Table \ref{tab:hpc}. While the number of \glspl{HPC} corresponds to the actual number of available counters, it is essential to note that the study's objective is to assess their effectiveness. It is plausible that by pinpointing a subset of \glspl{HPC}, acceptable classification results can still be achieved with a limited number of physical counter registers~\cite{Chenet:2025aa}.

Table \ref{tab:hpc} lists all selected logs, which include events that may not be available in standard RISC-V implementations, such as CVA6~\cite{cva6_doc}, due to their optional nature. We also provide x86 similarity data~\cite{intel_sdm}. Notably, the caches exhibit many similar events. Demand misses occur when the CPU requests an instruction not in the cache, requiring a fetch from lower memory levels. Read requests are a subset tracked when a read is issued to the gem5 cache component. This differentiation may not exist in real \glspl{HPC}, but our preprocessing reveals that the information content does not fully overlap.

\begin{figure*}[htb]
\centering
\begin{subfigure}{0.49\linewidth}
    \includegraphics[width=\linewidth]{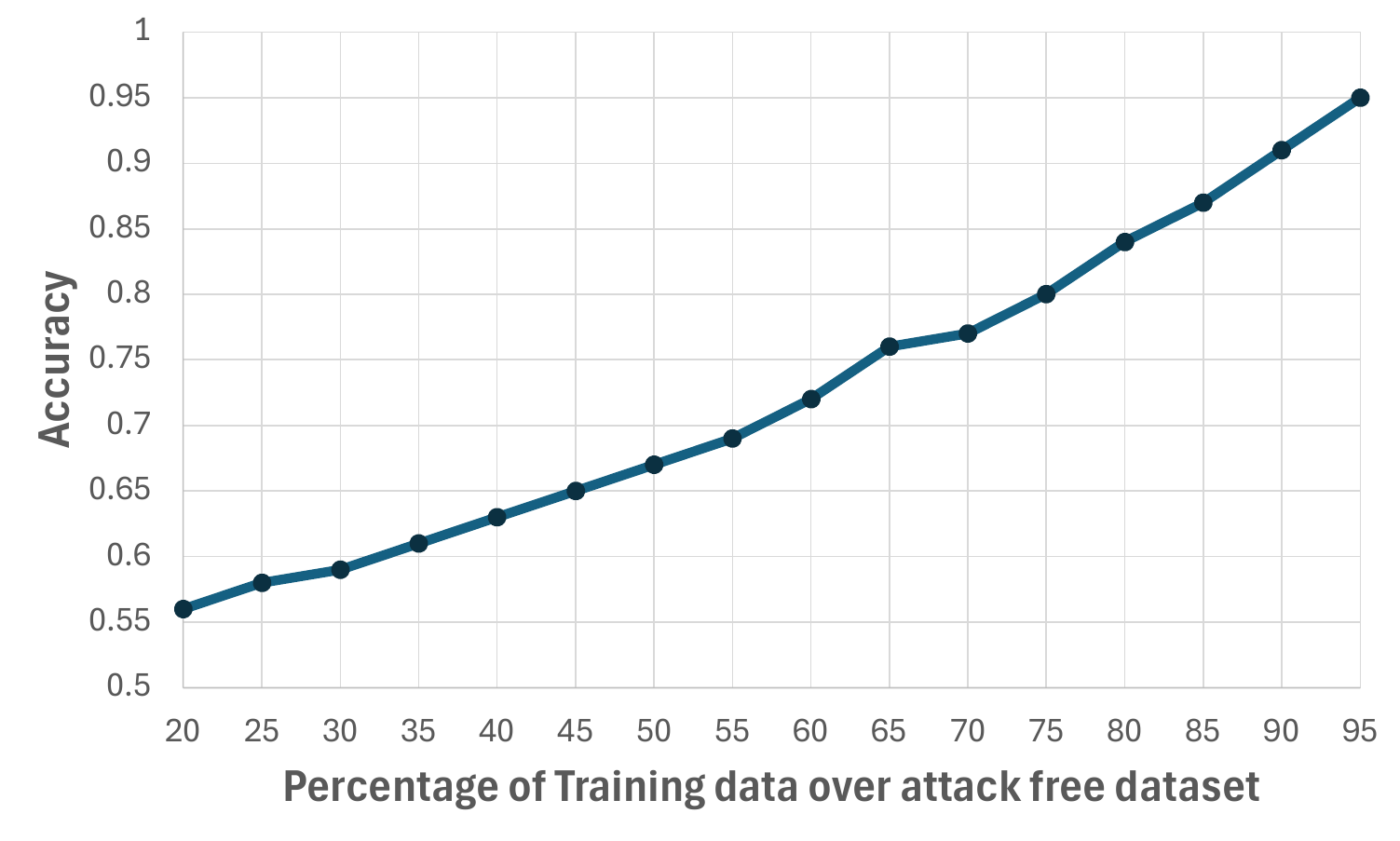}
    \caption{Attack detection Accuracy}
    \label{fig:dos_reducedAccuracy_test}
\end{subfigure}
\begin{subfigure}{0.49\linewidth}
    \includegraphics[width=\linewidth]{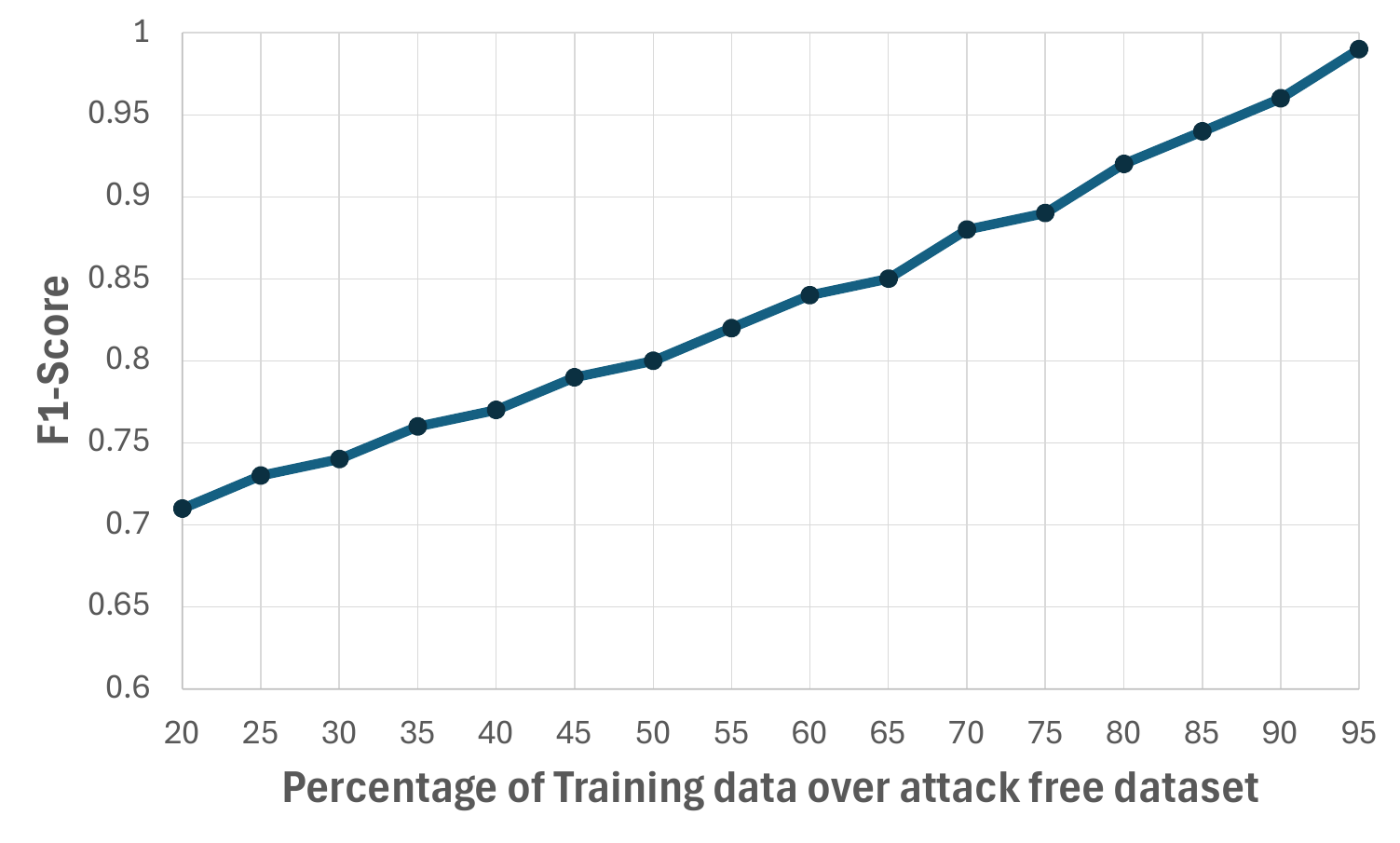}
    \caption{Attack detection F1-score}
    \label{fig:dos_reducedF1_test}
\end{subfigure}
\caption{\gls{IDS} performances on \gls{DoS} attack, considering a training set from 20\% to 95\% of the attack-free dataset in \cite{Hyunsung2017aa}}\label{fig:final_results_DoS_attack}
\end{figure*}

This study evaluates the effectiveness of one-class classification techniques in identifying various types of cyber attacks without prior knowledge of their characteristics. We focused on these methods' detection capabilities by varying the size of the attack-free training set, analyzing 1,500 \gls{CAN} frames processed by the \texttt{AES} task. Smaller training set sizes, starting with 75 frames, yielded inconclusive results and have been excluded from the final manuscript for clarity and integrity.

In Figure \ref{fig:final_results_DoS_attack}, we present essential performance metrics, specifically accuracy and the F1 score (see equation \ref{eq:f1}), as we vary the percentage of the dataset utilized for training. The accuracy metric assesses the ratio of correctly predicted instances to the total number of predictions made. Meanwhile, the F1 score offers a balanced evaluation of precision and recall, making it particularly valuable in situations with imbalanced class distributions.

\begin{equation}\label{eq:f1}
    F1 = \frac{TP}{{TP + \frac{1}{2}(FN + FP)}}
\end{equation}
The results show that the one-class classifier effectively identifies all malicious samples, but its performance heavily relies on the training set size. Achieving 90\% accuracy requires over 75\% of the training data (between 75\% and 80\% in Fig. \ref{fig:dos_reducedAccuracy_test}). This limitation means that without a sufficiently large and diverse training set, the model may misclassify legitimate traffic as malicious, as indicated by the F1-score in Fig. \ref{fig:dos_reducedF1_test}, which highlights the risk of high \gls{FPR} with inadequate training data.

Additionally, the classifier's ability to detect attacks depends on accumulating \gls{HPC} data over time, preventing instantaneous or real-time intrusion detection. This latency poses challenges for immediate threat response and concerns environments requiring prompt action against security breaches, which will be addressed in future work.

While the current classifier does not support real-time \gls{IDS} deployment, it demonstrates that malicious activities can cause significant deviations in application processing patterns. Understanding these deviations could enhance future detection algorithms and overall cybersecurity measures.

\section{Conclusion}
\label{sec:conclusion}

This paper introduced a novel \gls{IDS} approach utilizing \glspl{HPC} for detection. Initial findings demonstrate the feasibility of detecting attack \gls{CAN} data, although detection quality remains limited. The primary challenge is the need for a large dataset to train an effective model, indicating further experiments are necessary. Future work should explore various attacks, applications, and more complex \gls{RTOS} scenarios.

Additionally, research will focus on adapting \gls{IDS} capabilities for safety-critical real-time embedded systems, a vital step toward developing a comprehensive \gls{IDS}. The ultimate aim is to create an advanced \gls{IDS} that integrates \gls{CAN} bus anomaly detection and \gls{HIDS} into a unified module.

\section*{Acknowledments}
This paper is based on the work conducted by Gaspard Henri Guy Michel as part of their Master's thesis at Politecnico di Torino. Their dedication and research have significantly contributed to the findings and insights.

\bibliographystyle{IEEEtran}
\bibliography{biblio}

\end{document}